\documentclass[acmsmall,screen]{acmart}

\usepackage{fancyhdr}
\usepackage{listings}
\AtBeginDocument{%
  }

\setcopyright{none}
\copyrightyear{2024}
\acmYear{2024}
\acmDOI{}

\acmConference[ICFP miniKanren '24]{International Conference on Functional Programming, miniKanren and Relational Programming Workshop}{Sept 06, 2024}{Milan, Italy}
\begin{document}

\title{Improving stableKanren's Backward Compatibility}

\author{Xiangyu Guo}
\email{Xiangyu.Guo@asu.edu}
\orcid{0000-0001-8120-2365}
\affiliation{%
  \institution{Arizona State University}
  \streetaddress{699 S. Mil Ave}
  \city{Tempe}
  \state{Arizona}
  \country{USA}
  \postcode{85281}
}

\author{Ajay Bansal}
\email{Ajay.Bansal@asu.edu}
\orcid{0000-0001-8639-5813}
\affiliation{%
  \institution{Arizona State University}
  \streetaddress{699 S. Mil Ave}
  \city{Tempe}
  \state{Arizona}
  \country{USA}
  \postcode{85281}
}

\renewcommand{\shortauthors}{Xiangyu Guo and Ajay Bansal}

\settopmatter{printacmref=false}
\settopmatter{printfolios=true}
\renewcommand\footnotetextcopyrightpermission[1]{}
\pagestyle{fancy}
\fancyfoot{}
\fancyfoot[R]{miniKanren'24}
\fancypagestyle{firstfancy}{
  \fancyhead{}
  \fancyhead[R]{miniKanren'24}
  \fancyfoot{}
}
\makeatletter
\let\@authorsaddresses\@empty
\makeatother

\begin{abstract}
  We improve the backward compatibility of stableKanren to run miniKanren programs.
  stableKanren is a miniKanren extension capable of non-monotonic reasoning through stable model semantics.
  However, standard miniKanren programs that produce infinite results do not run as expected in stableKanren.
  According to stable model semantics, the contradictions are created by negations.
  A standard miniKanren’s relations do not involve negation, and the coarse contradictions handling in stableKanren causes this compatibility issue.
  Therefore, we provide a find-grinded contradiction handling to restrict the checking scope.
  As a result, standard miniKanren relations can produce answers.
  We also add a ``run-partial'' interface so that standard miniKanren’s relations implemented with ``define''/``defineo'' can generate answers even if they coexist with non-terminating or unsatisfiable stableKanren relations in the same environment.
  The ``run-partial'' interface also supports running stratified negation programs faster without checking global unavoidable contradictions.
  A dependency graph analysis can be applied to the input query in the future, so the ``run'' interface can implicitly decide whether to perform unavoidable contradictions checking to improve usability.
\end{abstract}




\received{10 June 2024}
\received[revised]{5 July 2024}
\received[accepted]{28 August 2024}

\maketitle
\thispagestyle{firstfancy}

\section{Introduction}
    The standard miniKanren introduces only a few operators: \textit{==} (\textit{unification}), \textit{fresh} (\emph{existential quantification}), \textit{conde} (\emph{disjunction}), and a \textit{run} (query) interface \cite{friedman2005reasoned}.
    stableKanren extends standard miniKanren to support non-monotonic reasoning through stable model semantics (Definition \ref{def:stable-model}) \cite{stableKanren}.
    In addition to the standard miniKanren operators, stableKanren adds two more operators: \textit{noto} (\emph{negation}) and \textit{defineo} (goal definition).

    Non-monotonic reasoning under stable model semantics resolves contradictions in the program, which can produce more than one model in some cases.
    Also, if the contradictions are unavoidable (Definition \ref{def:unavoidable-contradiction}) in non-monotonic reasoning, stable model semantics produce no model (unsatisfiable) for the program.
    For example, considering an unsatisfiable program in Listing \ref{lst:unsat}.
    \begin{lstlisting}[caption=An unsatisfiable program,label=lst:unsat]
(defineo (a) succeed) (defineo (b) succeed) (defineo (p) (a) (noto (p)))
    \end{lstlisting}
    % stableKanren did not distinguish fail and unsatisfiable results, it returns an empty list for both queries.
    % To verify the program is unsatisfiable, we need to query both positive and negative goals and get an empty list for both queries.
    There is a contradiction in \textit{p}, which requires \textit{p} to succeed and fail simultaneously.
    This contradiction is unavoidable, therefore stable model semantics consider the entire program to be unsatisfiable, even though the partial result can be obtained from \textit{a} and \textit{b}.

    In stableKanren, the resolution in \textit{run} interface continues checking the entire program even after obtaining the partial result for the query to ensure no unavoidable contradiction.
    However, stableKanren handles the checking very coarsely, which creates some issues for standard miniKanren programs.
    For example, consider the following miniKanren program without using \textit{noto} written in stableKanren (using \textit{defineo} instead of \textit{define}) in Listing \ref{lst:revo}.
    \begin{lstlisting}[caption=A ``revo'' relation to determine two lists are reversed to each other,label=lst:revo]
(defineo (revo xs sx)
  (fresh (empty) (nullo empty) (rev-acco xs empty sx)))
(defineo (rev-acco xs acc sx)
  (conde [(nullo xs) (== sx acc)]
         [(fresh (h t acc1)
            (conso h t xs) (conso h acc acc1) (rev-acco t acc1 sx))]))
    \end{lstlisting}
    The \textit{revo} determines if the two lists are reversed to each other.
    A simple query on the two empty lists does not terminate.
    \begin{lstlisting}
> (run 1 (q) (revo '() '()))
; infinity loop
    \end{lstlisting}
    If the relations in Listing \ref{lst:revo} are defined using \textit{define} instead of \textit{defineo}, the query can return a result.
    Also, the standard miniKanren relations implemented with \textit{define} fail to produce any answers if they coexist with unsatisfiable stableKanren relations (Listing \ref{lst:unsat}) in the same environment.

    We show in Section \ref{sec:normal-programs} that the unavoidable contradictions are introduced by negation (\textit{noto}), and the standard miniKanren programs do not include any negation are definite programs (Definition \ref{def:definite-program}).
    There is no need to check unavoidable contradictions for definite programs.
    Therefore, in Section \ref{sec:our-method}, we add a syntax analysis \textit{has-negation?} (Listing \ref{lst:has-negation}) to identify the definite program, and we modify stableKanren's \textit{defineo} (Listing \ref{lst:sk-defineo}) to improve the coarse unavoidable contradiction handling to a fine-grinded level (Listing \ref{lst:improved-sk-defineo}).
    Moreover, the unavoidable contradictions do not occur in the \textit{stratified negation programs} (Definition \ref{def:stratified-program}).
    The partial results in the \textit{stratified negation programs} are the final results.
    So, we recover the original miniKanren \textit{run} interface as \textit{run-partial} (Listing \ref{lst:run-partial}) in stableKanren.
    The only difference is that the \textit{run-partial} interface does not check unavoidable contradictions.
    We show in Section \ref{sec:result} that using \textit{run-partial} on final-SCC, a stratified program, is faster than \textit{run}.
    The \textit{run-partial} interface also allows the standard miniKanren relations implemented with \textit{define} to produce answers if they coexist with unsatisfiable stableKanren relations in the same environment.
    This different interface is beneficial for executing a single query, disregarding everything unreachable from it.
    In Section \ref{sec:conclusion}, we propose a dependency graph analysis that can be applied to the input query.
    So, the \textit{run} interface can implicitly decide to perform \textit{run-partial} or check unavoidable contradictions.

\section{Preliminaries}
    In this section, we review a few definitions, including definite programs (Definition \ref{def:definite-program}), normal programs (Definition \ref{def:normal-program}), stratified normal programs (Definition \ref{def:stratified-program}), and stable model semantics (Definition \ref{def:stable-model}).
    Then, we show a high-level overview of the outcomes of stable model semantics.
    Lastly, we review stableKanren and its coarse, unavoidable contradiction handling.
    
\subsection{Normal Programs and Stable Model Semantics}
\label{sec:normal-programs}
    Let us define normal programs.
    To begin with, we use the definition of \textit{definite program} and \textit{normal program} from John W. Lloyd \cite{DBLP:books/sp/Lloyd87}.

    \begin{definition}[definite program clause]
    A \textit{definite program clause} is a clause of the form,
    \[A \leftarrow B_1, \cdots, B_n\]
    where $A, B_1, \dots , B_n$ are atoms\footnote{Atom is evaluated to be true or false.}.
    \end{definition}
    A definite program clause contains precisely one atom A in its consequent.
    $A$ is called the \textit{head} and $B_1, \dots , B_n$ is called the \textit{body} of the program clause.
    \begin{definition}[definite program]
    \label{def:definite-program}
    A \textit{definite program} is a finite set of definite program clauses.
    \end{definition}
    
    Based on the definite program clause's definition, we have the definition for \textit{normal program clause} and \textit{normal program}.
    \begin{definition}[normal program clause]
    \label{def:normal-program-clause}
    A \textit{normal program clause} is a clause of the form,
    \[A \leftarrow B_1, \cdots , B_n, not \; B_{n+1}, \cdots , not \; B_{m}\]
    \end{definition}
    For a normal program clause, the body of a program clause is a conjunction of literals instead of atoms, $B_1, \cdots, B_n$ are \textit{positive literals} and $not \; B_{n+1}, \cdots, not \; B_{m}$ are \textit{negative literals}.
    \begin{definition}[normal program]
    \label{def:normal-program}
    A \textit{normal program} is a finite set of normal program clauses.
    \end{definition}

    According to Allen Van Gelder et al., there are sets of atoms named \emph{unfounded sets} in a normal program that can help us categorize the normal programs \cite{10.1145/116825.116838}.
    \begin{definition}[unfounded set]
    \label{def:unfounded-set}
    Given a normal program, the unfounded set is a set of atoms that only cyclically support each other, forming a loop.
    \end{definition}

    Considering the combinations of negations and unfounded sets (loops) in normal programs, we have informal definitions of \emph{stratified program}.
    \begin{definition}[stratified program]
    \label{def:stratified-program}
    Given a normal program, it is stratified if all unfounded sets (loops) do not contain any negation.
    \end{definition}

    The semantics of the definite programs and stratified programs are minimal model semantics \cite{Emden/Kowalski:1976:Semantics-of-Positive-LP}, and the semantics of the normal programs are stable model semantics \cite{Gelfond:1988:stable}.

    Michael Gelfond and Vladimir Lifschitz introduce \textit{stable model semantics} as the semantics for normal programs \cite{Gelfond:1988:stable}.
    Later, Miroslaw Truszczynski proposes an alternative reduct to the original definition: the alternative reduct leaves clause heads intact and only reduces clause bodies \cite{truszczynski2012connecting}.
    \begin{definition}[stable model semantics]
    \label{def:stable-model}
    Given an input program $P$, the first step is getting a \emph{propositional image} of $P$.
    A propositional image $\Pi$ is obtained from grounding each variable in $P$.
    The second step is enumerating all interpretations $I$ of $\Pi$.
    For a $\Pi$ that has $N$ atoms, we will have $2^N$ interpretations.
    The third step is using each model $M$ from $I$ to create a \emph{reduct program} $\Pi_M$ and verify $M$ is the minimal model of $\Pi_M$.
    To create a reduct program, we replace a negative literal $\lnot B_i$ in the clause with $\bot$ if $B_i \in M$; otherwise, we replace it with $\top$.
    Once completed, $\Pi_M$ is negation-free and has a unique minimal model $M'$.
    If $M = M'$, we say $M$ is a stable model of $P$.
    \end{definition}

    % Explain more about stable model semantics.
    Stable model semantics handle the loops containing negative literals in normal programs.
    There are two types of loops, odd and even, depending on the number of negative literals it has.
    Unlike minimal model semantics, which has only one model for a definite program and stratified program, stable model semantics can have three outcomes: no model, one model, or multiple models for a normal program.
    Overall, the number of negations in the loop controls the number of models.
    The even looped negation splits the models into more models.
    The odd looped negation creates a contradiction, and it must be avoided.
    There is no model if there is no way to bypass odd looped negation.

\subsection{stableKanren}
\label{sec:stable-model-semantics}

    stableKanren extends miniKanren under stable model semantics (Definition \ref{def:stable-model}) that support reasoning about contradiction \cite{stableKanren}.
    It defines a set of internal macros and functions that implicitly transform the positive goal from the user's input into a negative counterpart.
    Then, the positive and negative goals are unified under one goal function using \textit{defineo} macro.
    The \textit{negation counter} counts the number of \textit{noto} and decides to use a positive or negative goal during resolution.

    As a result, stableKanren can solve normal programs (Definition \ref{def:normal-program}). 
    For example, Maarten Herman Van Emden et al. introduce a two-person game \cite{10.5555/Two-PersonGames}.
    The game is given a directed connected graph, a peg on a starting node, and two players.
    Each player takes a turn to move the peg to the adjacency node; the winning move in this game is defined as there is a move that will make it so the opponent has no move.
    \begin{figure}[h]
    \centering
    \includegraphics[scale=0.5]{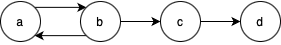}
    \caption{A playboard in two-person game}
    \label{fig:two-person-game}
    \Description{}
    \end{figure}
    Given an example graph in Figure \ref{fig:two-person-game} and the stableKanren program to find the winning positions as follows.

    \begin{lstlisting}
(defineo (move x y)
  (conde [(== x 'b) (== y 'c)] [(== x 'a) (== y 'b)]
         [(== x 'b) (== y 'a)] [(== x 'c) (== y 'd)]))
(defineo (win x) (fresh (y) (move x y) (noto (win y))))
    \end{lstlisting}
    In this example, the variable $x$ of \textit{win} unifies with values representing the winning positions.
    A set of queries and outputs through \textit{run} is shown as follows.
    \begin{lstlisting}
> (run 3 (q) (win q))
(*\textbf{(c b a)}*)
> (run 1 (q) (win 'c) (win 'a))
(*\textbf{(\_.0)}*)
> (run 1 (q) (win 'b) (win 'a))
(*\textbf{()}*)
    \end{lstlisting}
    The \textit{run} interface takes three parameters.
    The first parameter is the number of answers we expect; the query returns answers no more than this number.
    The second parameter is the query variable, which stores the answers found by the query.
    The third parameter is the actual query.
    The first query generates possible winning positions.
    The second query asks ``Can node \textit{c} and node \textit{a} both be the winning position?''
    It returns a list containing one element ``\_.0'', a representation of anything in miniKanren and stableKanren.
    Anything can let our queries succeed.
    So, \textit{a} and \textit{c} can both be the winning position.
    The last query asks: ``Can node \textit{b} and node \textit{a} both be the winning position?''
    It returns an empty list, representing nothing in miniKanren and stableKanren.
    In this case, nothing can let our queries succeed.
    Therefore, \textit{a} and \textit{b} can not be the winning position simultaneously.

    As Xiangyu Guo et al. mentioned, the partial result of the normal program (Definition \ref{def:normal-program}) may not be the final result since the odd looped negation may create an \emph{unavoidable contradiction} \cite{stableKanren}.
    \begin{definition}[unavoidable contradiction]
    \label{def:unavoidable-contradiction}
    An unavoidable contradiction means there is an atom in a normal program that can neither be proven true nor false because the odd negative loop always creates a contradiction.
    \end{definition}

    Recall the example program we showed in Listing \ref{lst:unsat}.
    An unsatisfiable result can be observed by running queries on the facts \textit{a} or \textit{b} and getting nothing.
    \begin{lstlisting}
> (run 1 (q) (a))        > (run 1 (q) (b))
(*\textbf{()}*)                       (*\textbf{()}*)
    \end{lstlisting}
    This shows the normal program does not have \textit{optimal substructure properties} and is categorized as NP-hard.
    Inside \textit{defineo} (Listing \ref{lst:sk-defineo}), the goal function adds to a checking set (\textit{program-rules}) for future unavoidable contradictions checking.
    %  Unavoidable Contradiction Handling
    \begin{lstlisting}[caption=A stableKanren defineo macro, label={lst:sk-defineo}]
(define-syntax defineo
  (syntax-rules ()
    ((_ (name params ...) exp ...)
      (begin
      (set! program-rules (adjoin-set (make-record `name 
                                        (length (list `params ...)))
                            program-rules))
      ;;; Omitted other details.
    ))))
    \end{lstlisting}
    However, not all goal functions are required to check for unavoidable contradictions.
    As we have shown in Section \ref{sec:normal-programs}, the definite program and stratified program are under minimal model semantics, and they have \textit{optimal substructure properties}.
    So, the partial result of these programs is also a part of the final result.
    As long as we can separate the definite programs from the normal programs in stableKanren, we improve the coarse contradiction checking to a fine-grinded level.

\section{Our Fix}
\label{sec:our-method}
    This section introduces a compilation time syntax analysis macro \textit{has-negation?} to identify definite programs.
    The standard miniKanren programs, like the one we see in Listing \ref{lst:revo}, are definite programs.
    \begin{lstlisting}[caption=A syntax analysis to identify negation in the program, label={lst:has-negation}]
(define-syntax has-negation?
  (syntax-rules (noto conde fresh)
    ((_ (noto g)) #t)
    ((_ (conde (g0 g ...) (g1 g^ ...) ...))
        (or (has-negation? g0 g ...)
            (has-negation? g1 g^ ...)
            ...))
    ((_ (fresh (x ...) g0 g ...))
        (has-negation? g0 g ...))
    ((_ g) #f)
    ((_ g0 g1 ...) 
        (or (has-negation? g0)
            (has-negation? g1)
            ...))))
    \end{lstlisting}
    The \textit{has-negation?} iterates through nested \textit{conde} and \textit{fresh} until it reaches the goal function level.
    If there is a negative goal function, it returns true, and vice versa.
    We apply this syntax analysis to prevent adding these goal functions to the checking set in Listing \ref{lst:improved-sk-defineo}.
    \begin{lstlisting}[caption=Improved defineo macro, label=lst:improved-sk-defineo]
(define-syntax defineo
  (syntax-rules ()
    ((_ (name params ...) exp ...)
      (begin
      (*\textbf{(if (has-negation? exp ...)}*)
        (set! program-rules (adjoin-set (make-record `name 
                                          (length (list `params ...)))
                              program-rules))(*\textbf{)}*)
      ;;; Omitted other details.
      ))))
    \end{lstlisting}
    The standard miniKanren program in Listing \ref{lst:revo} defined using our updated \textit{defineo} can produce a result.
    To run standard miniKanren programs with unsatisfiable stableKanren relations (Listing 1) in the same environment.
    We also recover the miniKanren \textit{run} as \textit{run-partial} in Listing \ref{lst:run-partial}.
    \begin{lstlisting}[caption=A run-partial interface without unavoidable contradictions checking, label=lst:run-partial]
(define-syntax run-partial
  (syntax-rules ()
    ((_ n (x) g0 g ...)
     (take n
       (lambdaf@ ()
         ((fresh (x) g0 g ... 
          (lambdag@ (negation-counter cfs c : S P)
              (cons (reify x S) '())))
          negation-counter call-frame-stack empty-c))))))
    \end{lstlisting}
    The \textit{run-partial} has the extended lambdag@ (adding negation counter and call stack frame) introduced by stableKanren but has no unavoidable contradictions checking.
    The \textit{run-partial} interface allows the standard miniKanren relations implemented with our updated \textit{defineo} to produce answers even if they coexist with unsatisfiable stableKanren relations (Listing \ref{lst:unsat}) in the same environment.

\section{Result}
\label{sec:result}
    As a result, all standard miniKanren relations defined by our updated \textit{defineo} can be run by stableKanren without any issues.
    The standard miniKanren relations can be combined with stableKanren relations in the same environment.
    If the \textit{run} interface produces unsatisfaction due to unavoidable contradictions, the user can still use the \textit{run-partial} interface to generate partial results without touching the contradictions part.
    The \textit{run-partial} interface can also run stratified programs without unavoidable contradiction checking.
    We demonstrate the performance gain over a stratified program using the final-SCC problem \cite{Warren1995ProgrammingIT}.
    \begin{figure}[ht]
    \centering
    \includegraphics[scale=0.5]{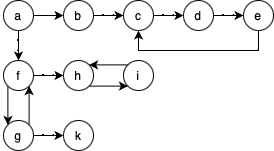}
    \caption{A directed graph contains multiple SCCs}
    \label{fig:finalSCC}
    \Description{}
    \end{figure}
    Consider the strongly connected components (SCCs) of Figure \ref{fig:finalSCC}.
    We will call an SCC a final SCC if the only nodes reachable from nodes in that SCC are others in that SCC.
    Now, given a node, we want to find all nodes that are reachable from that node in the final SCCs.
    The solution in stableKanren is as follows.
    \begin{lstlisting}
(defineo (edge x y)
  (conde
    [(== x 'a) (== y 'b)] [(== x 'b) (== y 'c)] [(== x 'c) (== y 'd)]
    [(== x 'd) (== y 'e)] [(== x 'e) (== y 'c)] [(== x 'a) (== y 'f)]
    [(== x 'f) (== y 'h)] [(== x 'f) (== y 'g)] [(== x 'g) (== y 'f)]
    [(== x 'g) (== y 'k)] [(== x 'h) (== y 'i)] [(== x 'i) (== y 'h)]))

(defineo (reachable x y)
  (conde [(edge x y)]
         [(fresh (z) (edge x z) (reachable z y))]))

(defineo (reducible x)
  (conde [(fresh (y) (reachable x y) (noto (reachable y x)))]))

(defineo (fullyReduce x y)
  (conde [(reachable x y) (noto (reducible y))]))
    \end{lstlisting}
    To run a query on \textit{fullyReduce} will produce all final-SCC pairs for us.
    \begin{lstlisting}
> (time (length (run-partial #f (q) (fresh (x y)
                          (fullyReduce x y) (== q `(,x ,y))))))
(time (length (run-partial #f ...)))
    2 collections
    0.013510916s elapsed cpu time, including 0.000435000s collecting
    0.013703000s elapsed real time, including 0.000441000s collecting
    11543824 bytes allocated, including 16744800 bytes reclaimed
100
> (time (length (run* (q) (fresh (x y)
                            (fullyReduce x y) (== q `(,x ,y))))))
(time (length (run* (...) ...)))
    5597 collections
    32.663375541s elapsed cpu time, including 4.063525000s collecting
    32.817000000s elapsed real time, including 4.092520000s collecting
    46932503280 bytes allocated, including 46927812840 bytes reclaimed
100
    \end{lstlisting}
    Getting the same 100 final-SCC pairs on the same machine, using \textit{run-partial} only takes 0.01 seconds, and using \textit{run} takes 32.81 seconds.
    If the user identifies their stableKanren program as a stratified program, they will benefit a huge performance gain from using \textit{run-partial} interface.

\section{Conclusion and Future Work}
\label{sec:conclusion}
    This paper presents an improvement to stableKanren's backward compatibility with miniKanren.
    We add compilation time syntax analysis \textit{has-negation?} to stableKanren's \textit{defineo} to achieve fine-grinded unavoidable contradictions checking.
    So, the definite programs (Definition \ref{def:definite-program}) defined by standard miniKanren relations without using \textit{noto} (negation) will not check for unavoidable contradictions.
    We also recover \textit{run-partial} execution time interface to allow standard miniKanren relations to coexist with stableKanren relations in the same environment.
    Therefore, the user can still produce partial results even if the programs have an unavoidable contradiction.
    Moreover, we show that running a stratified program in stableKanren using \textit{run-partial} is faster than \textit{run}.

    As a future work, the stratified programs can also be identified at compilation time using a macro like \textit{has-negation?} to analyze the dependency graph.
    So that one unique \textit{run} interface can implicitly decide to perform or not perform the unavoidable contradiction checking.
    We are working on this \textit{has-stratified?} macro to improve usability further.

% \begin{acks}

% \end{acks}

\bibliographystyle{ACM-Reference-Format}
\bibliography{ref}

\appendix

\end{document}